\documentclass{llncs}
\usepackage[T1]{fontenc}
\usepackage{graphicx}
\usepackage{hyperref}
\usepackage{color}

\usepackage[title]{appendix}

\usepackage{graphicx} 
\usepackage{booktabs} 
\usepackage{amssymb}
\usepackage{amsmath}
\usepackage{multirow}
\usepackage{tikz}
\usetikzlibrary{trees}
\usepackage{bbm}
\usepackage{bbding}
\providecommand{\Description}[2][]{}
\renewcommand{\Description}[2][]{}

\newcommand{\tool}{\textsc{ANVIL}}
\begin{document}
\title{\tool{}: Anomaly-based Vulnerability Identification without Labelled Training Data}
\titlerunning{\tool{}: Anomaly-based Vulnerability Identification w/o Labelled Data}
%
\author{Weizhou Wang\Envelope, Eric Liu, Xiangyu Guo, Xiao Hu, Ilya Grishchenko, David Lie
}
%
\authorrunning{W. Wang, E. Liu, X. Guo, X. Hu, I. Grishchenko and D.Lie}
%
\institute{University of Toronto, Toronto, Canada
\email{\{weizhou.wang,ec.liu,xiangyu.guo,hx.hu\}@mail.utoronto.ca}
\email{\{ilya.grishchenko,david.lie\}@utoronto.ca}}
%
\maketitle              

\newif\ifdraft
\draftfalse
\ifdraft
    \newcommand{\comment}[2]{{\textcolor{orange}{#1}}}
    \newcommand{\edit}[1]{{\textcolor{orange}{#1}}}
\else
    \newcommand{\comment}[1]{#1}
    \newcommand{\edit}[1]{#1}
\fi

\begin{abstract}
Supervised-learning-based vulnerability detectors often fall short due to limited labelled training data. In contrast, Large Language Models (LLMs) are trained on vast unlabelled code corpora, yet perform only marginally better than coin flips when directly prompted to detect vulnerabilities. In this paper, we introduce \tool{}, a line-level vulnerability localization approach that reframes vulnerability detection as anomaly detection, based on the premise that vulnerable code is rare and thus anomalous relative to patterns learned by LLMs. \tool{} performs a masked code reconstruction task: The LLM reconstructs a masked line of code, and deviations from the original are scored as anomalies. We propose a hybrid anomaly score that combines exact match, cross-entropy loss, prediction confidence, and structural complexity. We evaluate our approach across multiple LLM families, scoring methods, and context sizes, and against vulnerabilities after the LLM's training cut-off. On the PrimeVul dataset, \tool{} outperforms state-of-the-art supervised detectors -- LineVul, LineVD, and LLMAO -- achieving up to 2× higher Top-3 accuracy, 75\% better Normalized MFR, and a significant improvement on ROC-AUC. Finally, by integrating \tool{} with fuzzers, we uncover two previously unknown vulnerabilities, demonstrating the practical utility of anomaly-guided detection.
\end{abstract}

\section{Introduction}\label{sec:intro}
Probabilistic approaches to static vulnerability detection have garnered considerable research attention due to the promise of automatically detecting common bug patterns by training a model on historical data. Confidence in statistical modelling has only grown with recent advances in large language models (LLMs). One of the significant advantages of LLMs is that they can be trained in a self-supervised fashion without labelled data. This has enabled them to be trained on very large training sets, and led to impressive and surprising results in code understanding and generation~\cite{feng_codebert_2020,wang_codet5_2021}.

Unfortunately, the same cannot be said about their ability to detect vulnerabilities. As shown in a recent study~\cite{steenhoek_comprehensive_2024}, even with advanced LLMs like GPT-4~\cite{noauthor_chatgpt_nodate}, vulnerability prediction accuracy on real-world code is little better than a coin-flip. A key reason for this limitation is that LLMs are trained to generate code that resembles their training data, without explicit supervision about what constitutes buggy or vulnerable code. As a result, they lack the domain-specific knowledge needed to reliably distinguish vulnerable code from benign code. Meanwhile, existing learning and LLM-based vulnerability detectors, while attempting to embed the notion of vulnerabilities, still suffer from the lack of a large and well-labelled dataset~\cite{chen_diversevul_2023,ding_vulnerability_2024}. Efforts to compensate for the shortage of labelled data by selecting relevant code features (e.g., the granularity of the code context, commit history, control-flow, data-flow, etc.) and converting these features into a learnable representation (e.g., a vector or a graph)~\cite{li_vuldeepecker_2018,li_vulnerability_2021,cheng_how_2022,fu_linevul_2022,hin_linevd_2022,mirsky_vulchecker_2023,russell_automated_2018} have only yielded modest improvements. Similarly, recent attempts to generate new vulnerable training corpora~\cite{mirsky_vulchecker_2023,nong_vulgen_2023,nong_vgx_2024} have not improved real-world performance significantly, as synthetically generated vulnerabilities still cannot capture the huge variety of ways that vulnerabilities can manifest themselves.

LLMs are effective because they are trained on massive unlabelled datasets, enabling them to generate code in the same distribution. However, this generative capability does not translate to an ability to recognize vulnerable code, as such recognition is a different objective requiring additional training, ideally with labelled vulnerable code. Out of the huge amounts of unlabelled code, only a minuscule portion contains vulnerabilities, and even less has labels. 

Rather than trying to learn bug patterns from such sparse signals, we advocate for the inverse. Vulnerable code is more likely to fall outside the LLM's learned distribution, because vulnerabilities constitute only a small fraction of the code used in training. This allows us to recast bug detection as an \textit{anomaly detection} task: By comparing the original line of code to the generated version in a masked code reconstruction task, we determine whether the line is anomalous, i.e., out-of-distribution for an LLM.
This focus allows for isolating the vulnerability at a particular location, enabling line-level detection, while some vulnerability detectors only work at the function granularity~\cite{10.1145/3731557,li-etal-2025-clever,10.1145/3691620.3695013,wen_scale_2024,10.5555/3766078.3766104}.

There were attempts at modelling software defects as anomalies. For example, a study of code ``naturalness''~\cite{ray_naturalness_2016} leveraged statistical methods to model software bugs and their fixes, creating a purely statistical $10$-gram model trained on 35 million LOC for next-token prediction. Then, given a line of code, the model calculated the cross-entropy based on probabilities in its internal statistics and achieved a bug detection accuracy competitive with static code analyzers.

We hypothesize that the 10-gram model's result can be improved by modern transformer-based LLMs since they have considerably greater representational power, being trained on 10s-100s of billions of LOC (a 4-5 order of magnitude increase). However, it is not straightforward to adapt this previous work on code defects to vulnerability detection with LLMs. We highlight the following challenges: First, the entropy measurement used by previous work does not sufficiently discriminate between small text-level differences, which can be significant for software correctness. For instance, just a one-character difference in an if-statement's predicate can lead to an out-of-bound array access (e.g. > vs >=). To address this challenge, this paper explores different methods for quantifying the anomaly level of predictions. 
Second, unlike $10$-gram models, LLMs can use a context of hundreds to thousands of tokens to make predictions, raising the question of the appropriate context size to use, i.e., providing the LLMs with context that is both sufficient and yet not overly large so that it confuses the LLM. Third, the design space of LLMs includes diverse model architectures, training objectives and datasets. Thus, we also explore whether anomaly-based bug detection with LLMs generalizes to different model sizes and architectures. Finally, given that modern LLMs are trained on large datasets, there is a risk of ``leakage,'' that is, a pre-trained LLM may have seen both the vulnerable and fixed version of the same code. Therefore, we evaluate our approach on a leakage-free dataset containing only vulnerabilities disclosed after the LLMs’ training cutoff, demonstrating its ability to generalize to unseen vulnerabilities.

To summarize, the paper makes the following contributions:
\begin{itemize}
    \item We present \tool{}, a line-level anomaly-based vulnerability detection approach that does not require labelled training data. We demonstrate that \tool{}, leveraging pre-trained LLMs, effectively identifies vulnerable code as anomalous through masked code reconstruction.
    \item We evaluate different design choices, namely anomaly scoring functions, model selections and prompt contexts, and find that hybrid scoring, larger models, and compound statement contexts yield the best performance.
    \item \tool{} outperforms the state-of-the-art supervised-learning-based line-level detectors, namely, LineVul, LineVD, and LLMAO, with 1.8×–2× higher Top-3 accuracy, 40\%–75\% better Normalized MFR, and a significant improvement on ROC-AUC.
    \item We augment fuzzers with \tool{} to guide seed selection, uncovering two previously unknown vulnerabilities, one of which has been assigned a CVE.
\end{itemize}

We make all code for our tool ANVIL available at 
\url{https://github.com/joe-weizhou-wang/anvil-paper}.
All data and scripts used for running our evaluation are also made public for reproducibility.
\section{Methodology} \label{sec:main_idea}
\subsection{Overview} \label{sec:meth_overview}
\begin{figure*}[ht!]
  \centering
  \includegraphics[width=\linewidth]{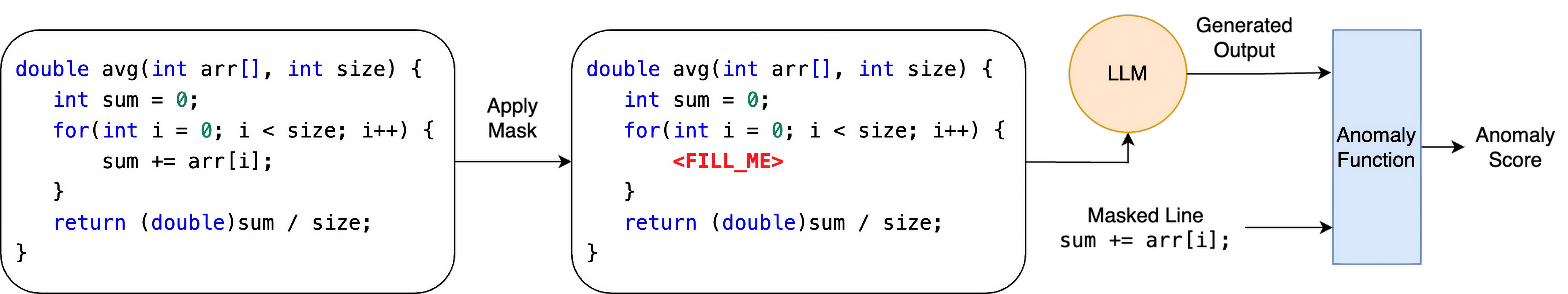}
  \caption{Overview of the \tool{}'s workflow}
  \Description[ROC-AUC for individual and hybrid scoring functions.]{
    The table reports ROC-AUC percentages for five scoring functions across context sizes from 2 to 500 lines. The hybrid score is best for nearly all context sizes, reaching 60.8 percent at 100 lines and 61.5 percent at 300 lines. The only exception is the 200-line context, where cross-entropy loss reaches 58.7 percent compared with 57.7 percent for the hybrid score.
    }
  \label{fig:system_block_diagram}
\end{figure*}

Similar to previous research~\cite{ray_naturalness_2016} in code ``naturalness,'' we consider a piece of code anomalous (i.e., unnatural) if a model considers it out-of-distribution. Figure~\ref{fig:system_block_diagram} describes the workflow of our technique, \tool{}. For each line of code under analysis, we mask (remove) the original (ground truth) line and instruct the LLM to reconstruct it based on the surrounding context. The LLM-generated code is then compared with the ground truth to compute an \textit{anomaly score}, which quantifies the divergence between them. 
LLMs are trained to predict the most probable token given a chain of previous tokens. Hence, a low reconstruction accuracy compared to the original code indicates that the original code is anomalous, i.e., has deviated from the learned representation of the model.
While our study focuses on C/C++ as a proof of concept, ANVIL is language-agnostic in design and can be applied to other programming languages supported by FIM–trained code LLMs.

\subsection{FIM \& Context Selection}\label{sec:FIM}
Code LLMs are trained for generative tasks such as code-completion, which predicts the next $n$ tokens given a code prefix, and fill-in-the-middle (FIM, or infilling)~\cite{bavarian_efficient_2022}, which generates tokens that fall between a given prefix and suffix. We focus on FIM for detecting anomalies. Although code completion is a more prevalent task, research by Bavarian et al.~\cite{bavarian_efficient_2022} highlights that FIM generally results in lower reconstruction loss due to additional constraints provided by the suffix. Suffix information allows FIM to capture downstream control and data flows, which are crucial for generating correct code for masked lines.

However, adapting FIM for anomaly detection introduces challenges. First, GPU memory and token limits make it infeasible to use entire source files as context.
Moreover, while LLMs claim to support long contexts, recent studies~\cite{liu-etal-2024-lost,li2025aixcoder7bv2trainingllmsfully,kate2025longfuncevalmeasuringeffectivenesslong} show that LLM performance often degrades when relevant information is buried within long contexts. 
As there is limited work on context selection for code reconstruction, let alone anomaly detection, we empirically evaluate two strategies: (1) a naive fixed-size approach and (2) a structure-aware approach. 

For (1), we investigate the effect of fixing the context to a specific number of source-code lines, equally split between the prefix and suffix. For instance, a 500-line context uses 250 lines preceding and succeeding the line under analysis. 
Naturally, this naive context selection may fail to capture relevant features beyond its fixed limits (e.g., a 100-line window applied to a 200-line function). Conversely, large contexts may pull in unrelated code, introducing noise that could confuse the model.

\begin{figure}[!h]
    \centering
    \includegraphics[width=0.7\linewidth]{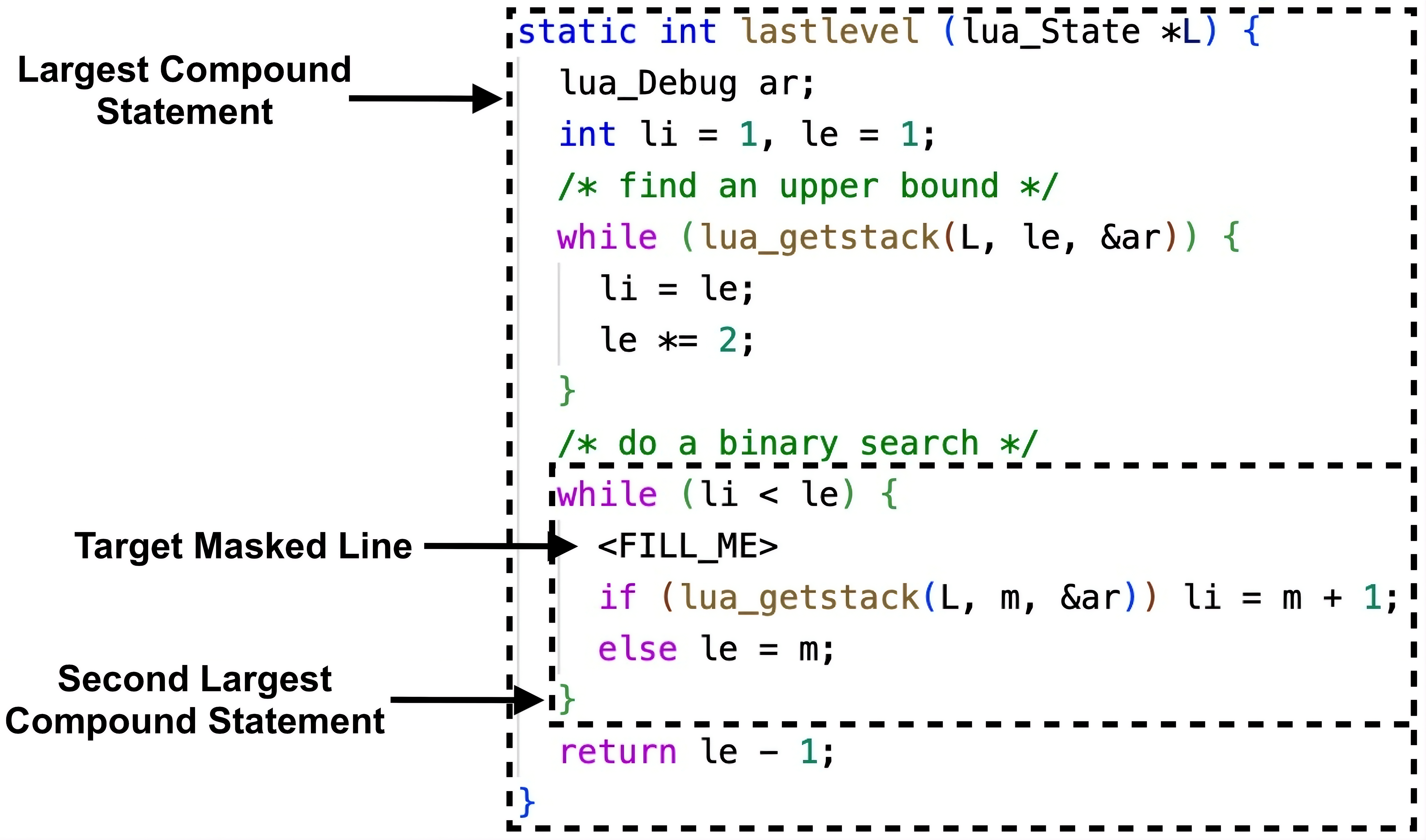}
    \caption{An example from \textit{lua} (a scripting language developed in C) showing the largest and second-largest compound statement for a line of interest}
    \Description[Compound-statement selection example.]{
    The figure shows a Lua C-code example used for adaptive context selection. Dashed boxes identify the largest compound statement enclosing the target masked line and the next smaller compound statement. Arrows label the largest compound statement, the target masked line represented by FILL_ME, and the second largest compound statement.
    }
    \label{fig:compstat_example}
\end{figure}

Alternatively, (2) uses an Adaptive Context (AC) approach, which selects the maximum compound statement enclosing the masked line.
This method ensures the model receives semantically relevant context without being distracted by irrelevant lines.
In C/C++, a compound statement refers to a sequence of statements enclosed by curly braces. Given a line of source code, we define the AC as the largest compound statement encapsulating the line. 
Often, the AC equates to the function body containing the line. However, if the AC is too large, exceeding token limits or GPU memory capacity, we resort to the next smaller compound statement that meets these constraints. We found that a context limit of 500 lines reached near-maximum peak usage of our GPU memory. Hence, our AC selection only considers compound statements that are 500 lines or fewer. Figure \ref{fig:compstat_example} demonstrates how we extract the compound statements associated with a masked line. 

\subsection{Anomaly Score}\label{sec:meth_anom_score}
We assign an \textit{anomaly score} to quantify the discrepancies between the generated line and the ground truth. We define the anomaly score as a function that takes the masked ground truth ($p$) and the generated line ($q$) as arguments and calculates a value $\delta(p,q)$, which estimates how ``anomalous'' the ground truth is, with higher values being more anomalous.

Previous work on code ``naturalness''~\cite{ray_naturalness_2016} uses cross-entropy loss for each token based on $10$-gram frequencies counted from their corpus to compute such an anomaly score. For LLMs, we compute the anomaly score as the average cross-entropy loss on the masked line, given its full surrounding context. This yields an overall loss measure $\delta_{loss}(p) = loss(p)$ for the line. 

However, loss values tend to be soft and diffuse, which could under-estimate anomaly severity: If a masked ground truth is ``$if (a >= 0)$'' and the model believes that the line is ``$if (a > 0)$'', the off-by-one-character ``='' might result in a low average loss but represents a potentially significant change, such as becoming the root cause of an out-of-bounds array access. Thus, we propose adding an exact match indicator as another metric, which treats the masked ground truth and the generated line as complete sequences. This approach assigns a value of 1 if the LLM's generated line matches the ground truth exactly, and 0 otherwise. An exact match implies that, during model inference for each token, the most probable token in each token distribution is perfectly aligned with the ground truth, and hence the line under analysis should not be considered anomalous. Thus, we can use the negation of exact match (EM) for anomaly scoring, $\delta_{EM}(p,q) = -\mathbbm{1}_{exact\_match}(p,q)$. Surprisingly, a crude measure like exact match has discriminative ability, as demonstrated by experiments in Section~\ref{sec:eval_anom_score}. 

\begin{figure}[!h]
    \centering
    \includegraphics[width=0.7\linewidth]{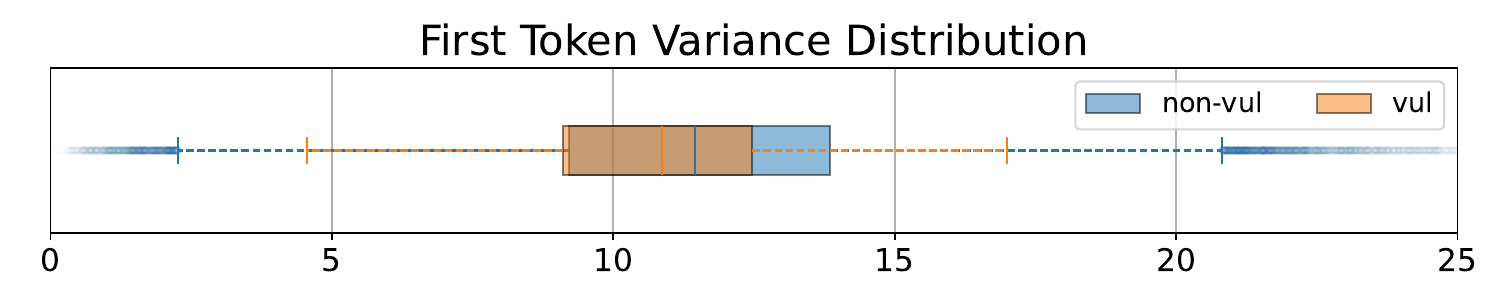}
    \caption{First generated token distribution variance on Magma dataset}
    \Description[First-token variance distributions for vulnerable and non-vulnerable lines.]{
    The figure is a horizontal box-plot comparison of first generated token variance on the Magma dataset. Non-vulnerable lines and vulnerable lines have overlapping central ranges around roughly 9 to 14, but non-vulnerable lines show a wider spread, including many very low and very high variance cases.
    }
    \label{fig:example_first_token_variance}
\end{figure}

Beyond the improved scale and ability of LLMs over the n-gram models used by previous work, LLMs expose more information for inferring model confidence. Namely, LLMs output probability distributions, which can be used to estimate their confidence in a prediction. We hypothesize that while there are perhaps only a few ways that code can be correct, there can be many ways that code can be incorrect, which should manifest as low-confidence predictions\footnote{We draw inspiration from Tolstoy who observed: ``All happy families are alike; each unhappy family is unhappy in its own way.''}. To estimate confidence, we observe differences in the models’ probability distributions for the first generated token when comparing vulnerable and benign lines. Specifically, we compute the variance among the top-10 predicted tokens at the first generation location, which reflects how concentrated or spread out the model's confidence is. Interestingly, as illustrated in Figure~\ref{fig:example_first_token_variance}, vulnerable lines tend to cluster around moderate variance, while benign lines exhibit a wider spread, mostly with very high and a few very low cases. High variance reflects strong model confidence, typical of benign lines with clear expected completions. In low variance cases, we find that the model is not necessarily ``confused,'' but is hedging between many valid completions, in syntactically simple code. This often arises in descriptive contexts (e.g., function or variable declarations), where there can be several plausible types or symbol names.

To better highlight such extremes, we transform the first-token variance into an anomaly score, as shown in Equation~\ref{scale_first_token_var}. 
Let $v$ denote the observed variance. The term $\overline{V}$, representing the empirical mean variance, serves as a calibration constant, while $C$ is a scaling constant. Both $\overline{V}$ and $C$ are determined empirically from a vulnerability dataset.
This transformation penalizes lines with variance near the average and favours those with more extreme values. 

\begin{equation}
    \label{scale_first_token_var}
    \delta_{var}(q) = -\exp(\frac{|v-\overline{V}|}{C})
\end{equation}
 
As noted above, there are syntactically simple cases, such as simple function invocations, where there can be many plausible benign completions. However, uncertainty about code can also be correlated with complex lines, such as nested conditionals or long expressions, where intricate logic may cause developers to overlook edge cases. To differentiate these two cases, we introduce an additional component based on the number of nodes in the Abstract Syntax Tree (AST) of the originally masked line, denoted $\delta_{ast}(p)$. A higher node count reflects greater lexical and syntactic complexity, which may signal higher vulnerability risk.

Our final anomaly scoring function is outlined in Equation~\ref{eq_anom_score}, aggregating all four components introduced above, each normalized.
This formulation rewards exact matches while capturing subtle deviations, model uncertainty, and code complexity between vulnerable and benign code. As shown in Section~\ref{sec:eval_anom_score} and Appendix~\ref{sec:ablation}, our experiments demonstrate that this combined approach consistently yields the most reliable vulnerability detection performance, with all components contributing positively to the overall effectiveness.

\begin{equation}
\label{eq_anom_score}
\delta_{\text{hybrid}}(p,q) 
=\, \delta_{\text{loss}}(p) + \delta_{\text{EM}}(p,q)
+\; \delta_{\text{var}}(q) + \delta_{\text{ast}}(p)
\end{equation}
\section{Evaluation}\label{sec:eval}

This section explores the effectiveness of \tool{}'s anomaly-based vulnerability detection through four research questions:
\begin{itemize}
    \item \textbf{RQ1:} How do different anomaly scoring methods affect \tool{}'s vulnerability detection capability?
    \item \textbf{RQ2:} Does an adaptive context size enhance vulnerability detection performance compared to a fixed context size?
    \item \textbf{RQ3:} How does \tool{} perform relative to supervised-learning-based vulnerability detection approaches?
    \item \textbf{RQ4:} Can \tool{}'s anomaly and vulnerability detection capabilities generalize to unseen data?
\end{itemize}

These questions investigate different aspects of \tool{}'s design and how varying parameters affect the efficacy of vulnerability detection. 
They also encompass comparisons between \tool{}, which requires no labelled inputs, and supervised vulnerability detectors, as well as an analysis of potential LLM data leakage.

\subsection{Experiment Setup}\label{sec:exp_setup}
All experiments were conducted using LLMs and tokenizers from the Huggingface library~\cite{wolf_transformers_2020}, on a machine equipped with an Nvidia H100 GPU with 80GB of HBM. For all LLMs, we used a temperature value of zero during code generation to prevent randomization.



We used three datasets to evaluate \tool{} across different dimensions. First, RQ1 and RQ2 explore how \tool{} performs under varying configurations, such as different scoring methods and context sizes. For these experiments, we use the Magma~\cite{hazimeh_magma_2020} benchmark due to its reliable vulnerability labelling. Specifically, Magma's labels are backed by a \textit{Proof of Vulnerability} (PoV) for each bug, representing an input that triggers the vulnerability. These PoVs are extracted through manual reviews and complemented by extensive fuzzing campaigns, ensuring that all labelled lines are directly tied to actual vulnerabilities. Magma contains 138 real-world CVEs from nine widely-used C/C++ projects (libpng, libtiff, libsndfile, libxml2, poppler, openssl, sqlite3, php, and lua), totalling 240,000 LoC and 5,227 functions. It includes both vulnerable and patched versions, with 256 lines labelled as vulnerable. Due to space limitations, we use CodeLlama-13B as the representative model for RQ1 and RQ2, as it is the most capable model in our evaluation. Comprehensive results covering all tested model architectures and sizes are included in Appendix~\ref{sec:model_arch_size}.

After identifying the most effective configurations of \tool{}, we compare it against various supervised-learning-based vulnerability detectors to address RQ3. To avoid potential overfitting from the previous configuration tuning process and to demonstrate \tool{}’s broader generalizability, the comparison is done on a different dataset, namely the PrimeVul~\cite{ding_vulnerability_2024} dataset, comprising over 700 open-source C/C++ projects. Despite its scale, PrimeVul maintains a low labelling error rate, achieving over 90\% accuracy via automated validation using descriptions from the NVD database~\cite{noauthor_nvd_nodate}.

Additionally, to address concerns of LLM data leakage, RQ4 evaluates \tool{} with vulnerabilities unseen by the LLMs during training. Because each LLM we tested is trained at a different time, it is difficult to find a large set of vulnerabilities that is guaranteed to be unseen by all models. Therefore, we focus on CodeLlama-13B for this leakage experiment because it generally provides the best performance as per Table~\ref{Tab:different_llm}. We leave the construction of leakage-free datasets and evaluation of other LLMs for future work. Using a script from the CVEFixes project~\cite{bhandari_cvefixes_2021}, we collected 100 CVEs from 53 repositories, all published between Dec 31, 2023, and Apr 20, 2024 (after CodeLlama-13B’s knowledge cutoff).  Instead of only relying on lines changed by defect-fixing commits, which may contain irrelevant information, we manually extracted only vulnerable lines from each fix. The final dataset includes 147 vulnerable functions (235 vulnerable and 10,956 non-vulnerable LoC) and 3,589 benign functions (101,173 LoC). We made this dataset publicly available to support future research.

In all experiments, the calibration factor ($\overline{V}$) and the normalization factors used in our anomaly function (Equation~\ref{eq_anom_score}) are derived from the Magma dataset. We also set the scaling constant $C$ to 40 because it consistently yields strong separation between vulnerable and benign lines on Magma. To avoid dataset-specific tuning, we keep all normalization factors fixed when applying the scoring function to other datasets.

\subsection{RQ1: Anomaly Score}\label{sec:eval_anom_score}
\begin{figure}[h!]
    \centering
    \includegraphics[width=0.95\linewidth]{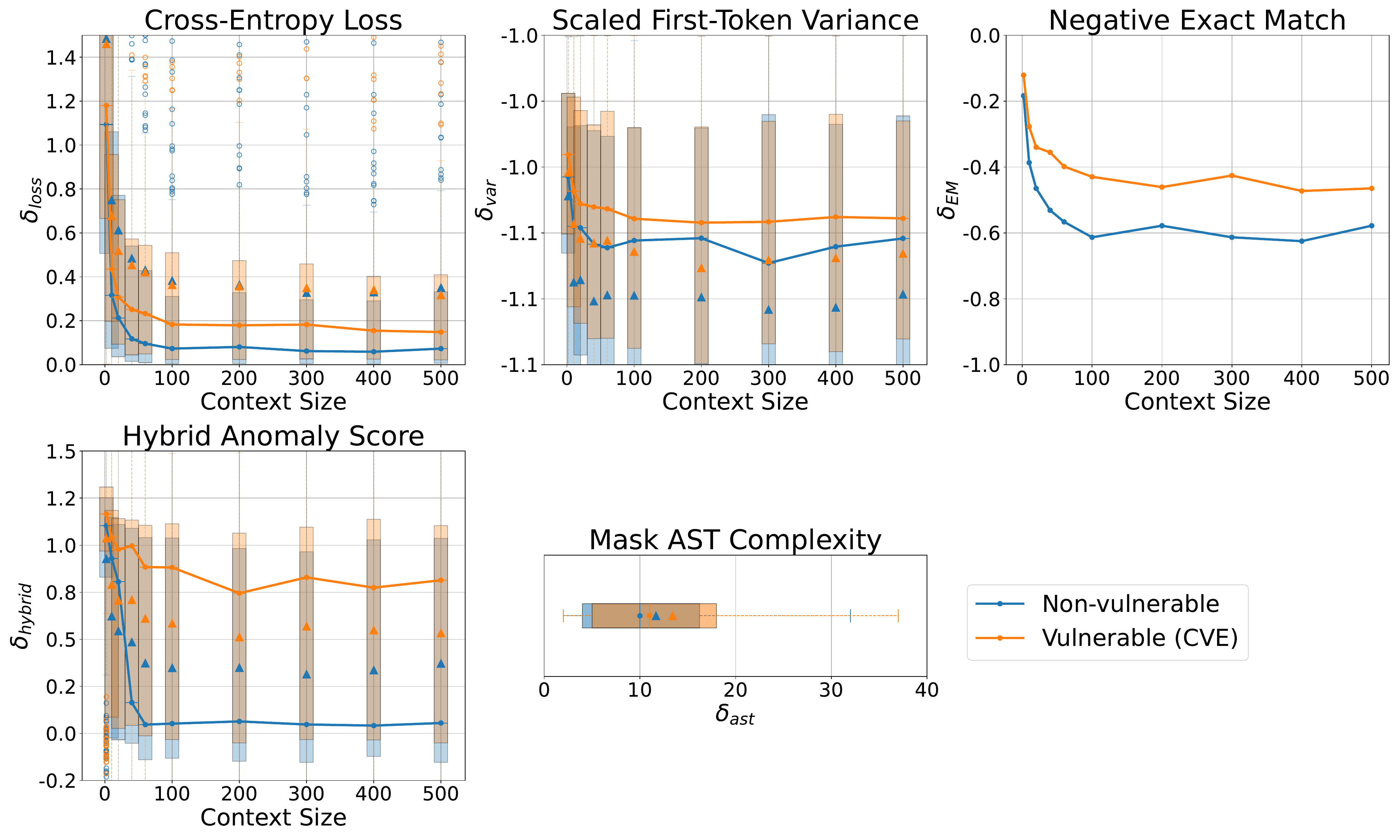}
    \caption{Anomaly scores on various functions}
    \Description[Anomaly-score components across context sizes.]{
    The figure compares vulnerable and non-vulnerable lines across five anomaly-score views: cross-entropy loss, scaled first-token variance, negative exact match, hybrid anomaly score, and masked-line AST complexity. Across context sizes, vulnerable lines generally have higher median cross-entropy loss and hybrid anomaly scores than non-vulnerable lines, while the hybrid score shows a clear separation that stabilizes for larger context sizes.
    }
    \label{fig:new_anomaly_score_2}
\end{figure}
To evaluate the effectiveness of the anomaly scoring function and its components (discussed in Section~\ref{sec:meth_anom_score}), we performed FIM (Section~\ref{sec:FIM}) using CodeLlama-13B on the Magma dataset. Specifically, we examine the score \emph{values} and the \emph{distributions} generated by each anomaly score for each line and evaluate the score's ability to discriminate between vulnerable and non-vulnerable lines. This evaluation was framed as a binary classification task, with the Area Under the Receiver Operating Characteristic Curve (ROC-AUC) used as a metric to measure the significance of the score in distinguishing between vulnerable and non-vulnerable lines. An ROC-AUC metric of 0.5 indicates random guessing, while higher values reflect a greater discrimination capability.

\begin{table*}[h!]
\centering
\Description[ROC-AUC for individual and hybrid scoring functions.]{
The table reports ROC-AUC percentages for five scoring functions across context sizes from 2 to 500 lines. The hybrid score is best for nearly all context sizes, reaching 60.8 percent at 100 lines and 61.5 percent at 300 lines. The only exception is the 200-line context, where cross-entropy loss reaches 58.7 percent compared with 57.7 percent for the hybrid score.
}
\caption{ROC-AUC using different scoring functions}
    \begin{tabular}{ |l|c|c|c|c|c|c|c|c|c|c|c|  }
         \hline
         & \multicolumn{9}{c|}{Context Size} &\\
         \hline
         Func & 2 & 10 & 20 & 40 & 60 & 100 & 200 & 300 & 400 & 500\\
         \hline
         $\delta_{ast}$ & 55.6\% & 55.6\% & 55.6\% & 55.6\% & 55.6\% & 55.6\% & 55.6\% & 55.6\% & 55.6\% & 55.6\% \\
         $\delta_{EM}$ & 53.1\% & 55.5\% & 56.2\% & 58.8\% & 58.4\% & 59.2\% & 55.9\% & 59.4\% & 57.6\% & 55.7\%\\
         $\delta_{var}$ & 54.5\% & 57.6\% & 54.5\% & 56.4\% & 56.9\% & 55.1\% & 52.9\% & 54.5\% & 55.3\% & 53.9\%\\
         $\delta_{loss}$ & 51.8\% & 54.3\% & 53.8\% & 57.2\% & 57.8\% & 58.1\% & \textbf{58.7\%} & 60.4\% & 59.5\% & 57.6\%\\
         $\delta_{hybrid}$ & \textbf{56.9\%} & \textbf{57.8\%} & \textbf{57.2\%} & \textbf{60.7\%} & \textbf{60.7\%} & \textbf{60.8\%} & 57.7\% & \textbf{61.5\%} & \textbf{60.1\%} & \textbf{57.7\%}\\
         \hline
    \end{tabular}
    \label{Tab:scoring_function}
\end{table*}

To compute the values of each scoring function component in FIM, we used all 256 vulnerable LOC from the Magma dataset as vulnerable samples. To equally reward vulnerable and non-vulnerable lines in the comparative analysis, we balanced the dataset by randomly sampling 256 non-vulnerable lines (excluding comments and whitespace).
For each reconstructed line of code, we calculate anomaly scores using five scoring functions (described in Section~\ref{sec:meth_anom_score}) across various fixed context sizes from 2 to 500 lines. Figure~\ref{fig:new_anomaly_score_2} presents the values of each component. For $\delta_{EM}$, we report the averaged values due to their binary nature. 
For the $\delta_{loss}$, $\delta_{var}$, and $\delta_{hybrid}$ scores, we plot median values as connected lines across context sizes to reduce the influence of outliers.
To provide a comprehensive view, we also included box plots~\cite{noauthor_box_2025}, with mean values represented by small triangles. Additionally, since $\delta_{ast}$ is a static property of the masked line itself and independent of LLM output, we present its distribution only as a box plot.

Across all scoring function components, the median values reveal a noticeable gap between categories of vulnerable and non-vulnerable lines, and this gap typically stabilizes when the context size exceeds 100 lines. This observation reinforces the notion that the LLM processes these two categories differently. Conversely, as the context size decreases below 100 lines, the anomaly scores for both vulnerable and non-vulnerable lines gradually increase, and the gap between the two categories narrows, indicating that the LLM becomes less accurate in reconstructing the masked lines regardless of their categories. This performance decline occurs because smaller context sizes do not provide the model with sufficient information for accurate generation.
Among the scoring functions, the $\delta_{hybrid}$ function exhibited a larger median gap between vulnerable and non-vulnerable lines compared to other scoring functions.

To quantify these findings, we computed ROC-AUC for the distributions of vulnerable and non-vulnerable lines for each scoring function component and context size pair, as shown in Table~\ref{Tab:scoring_function}. For each context size, we highlighted in bold the component with the highest ROC-AUC. The results indicate that the combined scoring function ($\delta_{hybrid}$, Equation~\ref{eq_anom_score}) yields the highest ROC-AUC on nearly all context sizes (with statistical significance $p < 0.0002$, using the method of DeLong et al.~\cite{65f9f828-9f33-36dc-9429-5d215792ea89}), with only a single exception where $\delta_{loss}$ slightly outperforms it on a context size of 200 lines. 
Further computing \textit{Spearman’s Rho Correlation} (SRC) and \textit{Cohen’s d} between each pair of components within $\delta_{hybrid}$ reveals strong separation across all pairs (SRC $< 0.31$ and d $> 1.7$, except $\delta_{loss}$ and $\delta_{EM}$), highlighting that every component makes valuable contribution to $\delta_{hybrid}$ from a distinct perspective. We further validate these observations through an ablation study in Appendix~\ref{sec:ablation}, which demonstrates that removing any component of $\delta_{hybrid}$ consistently degrades performance.

\subsection{RQ2: Adaptive Context Size}\label{sec:eval_mcs}

\begin{figure}[h!]
    \centering
    \includegraphics[width=1\linewidth]{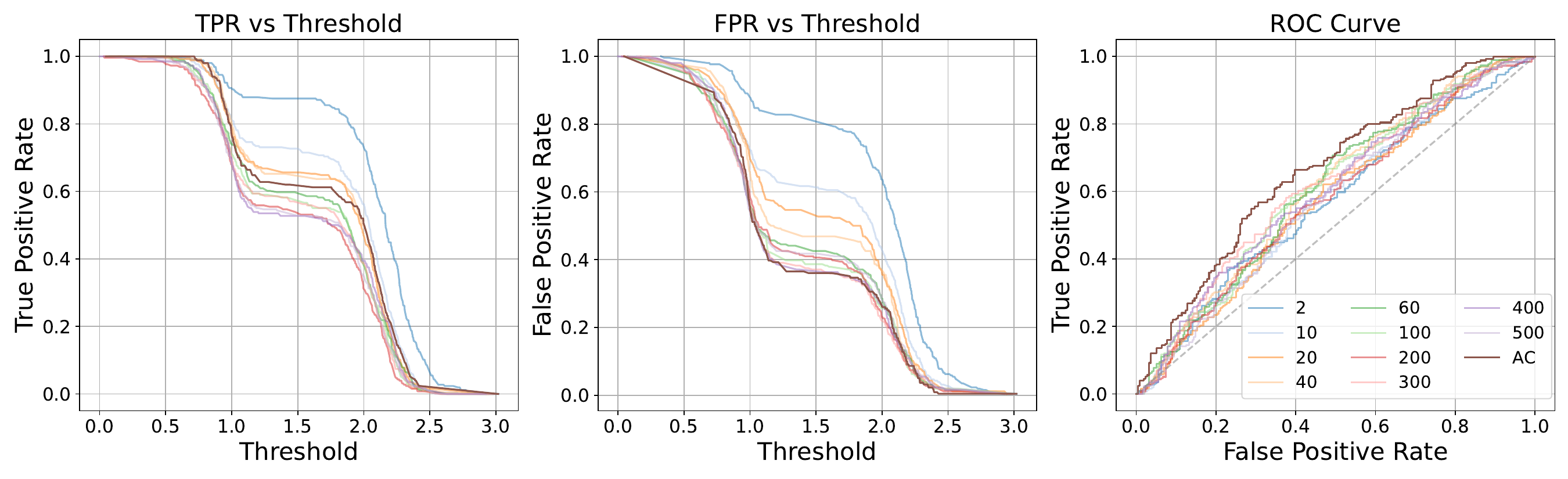}
    \caption{TPR, FPR \& ROC under different contexts}
    \Description[TPR, FPR, and ROC curves for fixed and adaptive contexts.]{
    The figure contains three line charts comparing fixed context sizes and adaptive context. The true-positive-rate and false-positive-rate curves vary by threshold, and the ROC plot shows adaptive context above the fixed-context curves for much of the range, indicating better discrimination between vulnerable and non-vulnerable lines.
    }
    \label{fig:roc_curve}
\end{figure}

As shown in Figure~\ref{fig:new_anomaly_score_2}, varying fixed context sizes affect average discrimination ability. However, static windows may confuse the model by including irrelevant or omitting critical code. This leads to fluctuations in performance and makes it difficult to identify an optimal context size. 

To address this limitation, we propose Adaptive Context (AC) as defined in Section~\ref{sec:FIM} -- a structure-aware approach that adaptively selects meaningful context for each line, aiming to provide a more stable and accurate anomaly score.
We evaluate AC using the CodeLlama-13B model with the $\delta_{hybrid}$ score and compare it to fixed context sizes.
Adopting AC significantly improves the ROC-AUC to 66.3\%\footnote{Running on the full Magma dataset yielded a consistent ROC-AUC of 65.8\%, confirming our sampling introduced no significant bias.}, outperforming all fixed-size settings presented in Table~\ref{Tab:scoring_function} (with $p < 0.0001$ on all pairs). A deeper investigation (Figure~\ref{fig:roc_curve}) reveals that AC effectively reduces the false positive rate compared to fixed-size contexts, likely due to its ability to focus only on relevant context and avoid the lost-in-the-middle issue of longer inputs.
Additionally, when using fixed contexts, we encountered three samples that triggered CUDA out-of-memory (OOM) errors at large sizes (400 and 500 lines) due to long line lengths. In contrast, AC avoided these OOMs and reduced peak memory usage to 67GB.

\begin{figure}[h!]
    \centering
    \includegraphics[width=0.5\linewidth]{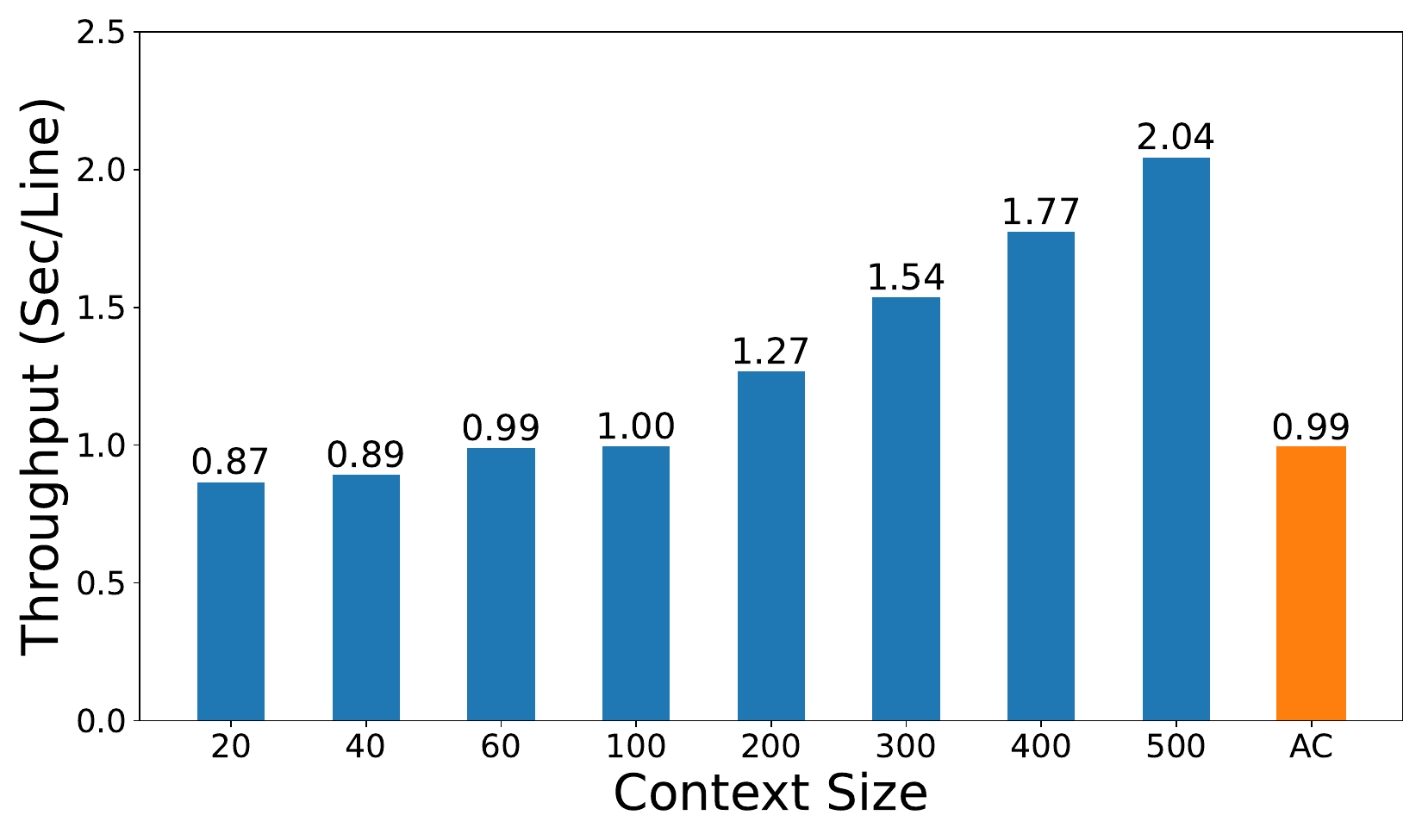}
    \Description[Runtime per line for fixed and adaptive contexts.]{
    The bar chart reports inference time in seconds per line for fixed context sizes and adaptive context. Runtime increases as fixed context grows, from 0.87 seconds at 20 lines to 2.04 seconds at 500 lines. Adaptive context takes 0.99 seconds per line, similar to the 60-line and 100-line fixed contexts and much faster than large fixed contexts.
    }
    \caption{Throughput (Sec/Line) on fixed context sizes vs. adaptive context}
    \label{fig:fixed_vs_msc_roc}
\end{figure}

We further measured the time required to run experiments across different context sizes. As shown in Figure~\ref{fig:fixed_vs_msc_roc}, inference time increases with larger fixed windows, with the 500-line context taking over 2 seconds to analyze a line -- twice as long as the 100-line window. This slowdown is due to the greater computational demands of larger contexts, which increase hardware contention and reduce throughput. In contrast, using AC significantly reduced runtime, making it, on average, nearly as fast as the 100-line fixed window. To understand this, we analyzed the distribution of AC sizes across all samples and found that most were relatively short (median of 86.5 lines).

To summarize, AC dynamically selects semantically meaningful code blocks. By mitigating the limitations of fixed contexts, it achieves the best discrimination between vulnerable and non-vulnerable lines while improving computational efficiency by lowering GPU memory usage and speeding up inference.

\subsection{RQ3: Line-level Vulnerability Detection}\label{eval:vul_detect}

We compare \tool{}, which does not require any labelled data, against state-of-the-art supervised line-level vulnerability detectors: LineVul, LineVD, and LLMAO. These baselines were selected because they operate at the same line-level granularity as \tool{}, ensuring a fair and consistent comparison. LineVul~\cite{fu_linevul_2022} and LineVD~\cite{hin_linevd_2022} both fine-tune CodeBERT~\cite{feng_codebert_2020} on the BigVul~\cite{fan_cc_2020} dataset; LineVul uses attention scores from transformer layers, while LineVD incorporates Joern's graph representations~\cite{noauthor_joern_nodate}. 
LLMAO~\cite{yang_large_2024} is an LLM-based line-level detector built on CodeGen~\cite{nijkamp_codegen_2023}, using models up to 16B parameters and a fixed 128-line context window. It was originally trained on Devign~\cite{zhou_devign_2019}, which includes only two projects (FFmpeg and QEMU). When evaluated using their released checkpoint, LLMAO exhibits substantially lower performance on PrimeVul than reported in the original study, likely due to the limited coverage of its original training set (Devign). To ensure consistency with other baselines, we retrained LLMAO on BigVul, which covers 4,296 projects. While retraining improves performance across all metrics (Table~\ref{Tab:primevul_baseline_cmp}, last row), the resulting ROC-AUC still remains below the 0.855 reported by Yang et al.

For experiments, we configured \tool{} to use the CodeLlama-13B model, the hybrid anomaly score (Equation~\ref{eq_anom_score}), and Adaptive Context (AC), as this configuration generally yields the best performance (see previous RQs). We evaluate \tool{} using the same two types of metrics employed by our baselines.
\begin{enumerate}
    \item We approach vulnerability detection as a binary classification task and measure the line-level classification performance using ROC-AUC. 
    This metric treats each line of code as an independent data sample, assessing the overall classification accuracy without setting a specific threshold.
    \item To gauge how effective \tool{}'s anomaly score is as a vulnerability prioritization tool, we evaluate all detectors with a ranking task. We use Top-N accuracy~\cite{noauthor_top_k_accuracy_score_nodate} (a standard metric for measuring vulnerability prioritization ability) to assess whether any of the vulnerable lines are included in the detector’s top-n most confident predictions within each function. Additionally, we report the \textit{Mean First Ranking} (MFR) to provide an overall measure. 
\end{enumerate}

We performed evaluations on the PrimeVul dataset described in Section~\ref{sec:exp_setup}. We selected this dataset over Magma and BigVul to mitigate potential overfitting: The supervised-learning-based detectors were trained and tuned on the BigVul dataset, whereas \tool{} does not require labelled data; however, it uses calibration and scaling constants V and C, as well as normalization factors, that are derived from Magma (introduced in Section~\ref{sec:meth_anom_score}). These processes could introduce biases specific to their respective datasets. By using PrimeVul as a separate evaluation dataset, we ensure independence and promote a fairer comparison.

\begin{table}[t!]
\centering
 \Description[PrimeVul vulnerability detection and localization results.]{
The table compares ANVIL with LLMAO, LineVD, and LineVul on PrimeVul using Top-1, Top-3, Top-5, MFR, normalized MFR, and ROC-AUC. ANVIL has the best Top-1, Top-3, Top-5, normalized MFR, and ROC-AUC values: 11.7 percent, 31.3 percent, 41.4 percent, 0.18, and 61.8 percent. LineVul has the lowest raw MFR, but performs worst after normalization.
}
\caption{Vulnerability Detection \& Localization on PrimeVul Dataset}
    \begin{tabular}{@{\hspace{0.5em}}l@{\hspace{0.5em}}l@{\hspace{0.5em}}l@{\hspace{0.5em}}l@{\hspace{0.5em}}l@{\hspace{0.5em}}l@{\hspace{0.5em}}l@{\hspace{0.5em}}l} \toprule
        {Method} & {Dataset} & {Top-1$\uparrow$} & {Top-3$\uparrow$} & {Top-5$\uparrow$} & {MFR$\downarrow$} & {N-MFR$\downarrow$} & {ROC-AUC$\uparrow$} \\ \midrule
        ANVIL & N/A & \textbf{11.7\%} & \textbf{31.3\%} & \textbf{41.4\%}  & 25.0 & \textbf{0.18} & \textbf{61.8\%} \\ \midrule
        LLMAO & BigVul & 9.2\% & 17.3\% & 26.2\% & 49.3 & 0.39 & 51.8\% \\ \midrule
        LineVD & BigVul & 9.2\%  & 15.4\% & 23.1\% & 38.0 & 0.30  & 49.6\% \\ \midrule
        LineVul & BigVul & 4.0\% & 17.0\% & 30.0\% & \textbf{17.6} & 0.71 & 49.6\% \\ \midrule \midrule
        LLMAO & Devign & 4.4\% & 14.0\% & 22.5\% & 50.7 & 0.40 & 50.7\% \\ \bottomrule
    \end{tabular}
    \label{Tab:primevul_baseline_cmp} \\
\end{table}

\subsubsection{Line-level Vulnerability Classification.}

We first evaluate \tool{} against LineVul, LineVD, and LLMAO in their ability to classify lines with vulnerabilities using ROC-AUC scores. This experiment uses the PrimeVul dataset with several accommodations to ensure fairness and compatibility. First, following the \textit{independent testing} principle outlined in \cite{nong_vgx_2024,nong_vulgen_2023,chen_diversevul_2023}, we ensured that the testing and training sets include data from different sources to avoid bias caused by duplicated data distributions. Specifically, we included only projects not part of the training datasets (BigVul and Devign) used for the baselines. For the remaining data, we included only functions with corresponding source files available to support varying context window sizes used in \tool{} and LLMAO. Furthermore, in line with LineVD and LLMAO, all whitespace lines were removed from the source files. We also considered only vulnerable lines with paired patched versions. Comments were then filtered out to create the final versions of vulnerable lines. This process resulted in a dataset containing 924 vulnerable LoC and 33,820 non-vulnerable LoC from 273 vulnerable functions, along with 325,353 LoC from 9,273 benign functions. These vulnerabilities span 40 different CWEs, with detailed breakdowns provided in~\cite{appendix_table}.

Additionally, as \tool{} operates differently from the three other tools, we made the following adjustments. Due to the way LineVul uses CodeBERT (processing entire functions as single inputs), 
it inherits CodeBERT’s 512-token limit. For functions exceeding this limit, LineVul truncates them to include only the first 512 tokens, discarding the rest. Consequently, we evaluate LineVul only on lines it retains, which is 455 vulnerable LoC and 192,382 non-vulnerable LoC in the dataset, whereas the other three tools are evaluated on the full dataset of approximately 360k LoC. 

As shown in Table~\ref{Tab:primevul_baseline_cmp}, \tool{} achieves ROC-AUC scores of 61.8\% with CodeLlama-13B, outperforming all three baselines (with $p<0.0001$). 
In contrast, LineVul, LineVD and LLMAO yield only ROC-AUC scores close to that of random guessing (approximately 50\%). Although these results are lower than the original performance reported in the respective papers, they align with findings from the DiverseVul study~\cite{chen_diversevul_2023}, which demonstrated that supervised vulnerability detectors experience significant performance degradation when evaluated on projects not seen during training. This observation has been further validated by subsequent works~\cite{nong_vgx_2024,ding_vulnerability_2024}. In contrast, since \tool{} does not rely on labelled training data, its anomaly-based detection strategy exhibits stronger generalization to different projects.

As ROC-AUC captures overall detection performance across all thresholds, we use it as our primary metric; Precision, Recall, and F1 scores are reported in Appendix~\ref{sec:rq3_PRF1} for completeness.

\subsubsection{Vulnerability Prioritization.} 
This experiment evaluates the ability of vulnerability detectors to prioritize vulnerable lines within a given function by comparing their Top-N accuracies and Mean First Ranking (MFR). Specifically, Top-N accuracy measures whether at least one vulnerable line is included among the top N lines ranked by each detector. For \tool{}, lines are ranked based on anomaly scores, while LineVul ranks lines using attention scores, and LineVD and LLMAO rely on logits from their respective neural network heads. We report Top-1, Top-3, and Top-5 accuracies, as these metrics are originally used by the baselines. Additionally, we calculate MFR (smaller is better) by averaging the rank of 
the highest-ranked vulnerable line 
in each function, providing a comprehensive measure of overall vulnerability prioritization performance. This experiment uses 273 vulnerable functions from the PrimeVul dataset, consisting of 924 vulnerable LoC and 33,820 non-vulnerable LoC.

As shown in Table~\ref{Tab:primevul_baseline_cmp}, \tool{} outperforms all baselines in Top-N accuracy. While it achieves a modest 1.27 times improvement in Top-1 accuracy over the best-performing baseline, LLMAO, its advantage becomes more pronounced when the criterion is relaxed to Top-3 and Top-5, with \tool{} achieving 1.81 and 1.58 times higher accuracies, respectively. Notably, the 512-token limit of LineVul results in an average of only 25 available LoC per function, compared to an average function length of 141 LoC for the other detectors. This shortened context theoretically gives LineVul an advantage in achieving higher Top-N accuracy due to the smaller candidate pool. However, its performance falls short, achieving only 4.0\% Top-1 accuracy (the lowest among all detectors), and its Top-3 and Top-5 accuracies remain below those of \tool{}.

The limited function length also explains why LineVul appears to achieve the best MFR. To enable a fairer comparison, we use Normalized Mean First Ranking (N-MFR), which adjusts the raw MFR based on average function length. For LineVul, we apply a normalization factor of 25 LoC, while for all other detectors, we use an average function length of 141 LoC. The normalized results show that \tool{} achieves the best N-MFR of 0.18, which is 54\% lower than LLMAO and 40\% lower than LineVD. Although LineVul originally records the lowest raw MFR, its normalized score rises to 0.71, 
yielding the worst N-MFR score.

\subsection{RQ4: Data Leakage}\label{sec:eval_leakage}
In this section, we address concerns about potential data leakage that may arise if patches for the vulnerabilities in our evaluation dataset were present in the LLM training data, as this may enable the LLM to correctly generate the patched code instead of the vulnerable code, as discussed in Section~\ref{sec:exp_setup}. The evaluation is conducted on CodeLlama-13B with the hybrid scoring function and Adaptive Context.

We assess the generalizability of \tool{} in detecting line-level vulnerabilities by comparing its detection performance on PrimeVul and on our newly collected 2024 CVEFixes dataset described in Section~\ref{sec:exp_setup}. Similar to RQ3, this evaluation examines both vulnerable line identification and prioritization performance.

\begin{table}[!h]
\centering
\Description[Generalization results on PrimeVul and 2024 CVEFixes.]{
The table compares ANVIL on PrimeVul and CVEFixes. CVEFixes has higher Top-1, Top-3, Top-5, and ROC values, with 16.4 percent, 33.6 percent, 47.3 percent, and 68.6 percent, compared with 11.7 percent, 31.3 percent, 41.4 percent, and 61.8 percent on PrimeVul. Normalized MFR is similar across datasets, 0.18 on PrimeVul and 0.22 on CVEFixes.
}
\caption{\tool{}'s generalizability on 2024 CVEFixes Dataset}
    \begin{tabular}{llllllll} \toprule
        {Tested On} & {Top-1} & {Top-3} & {Top-5} & {MFR} & {N-MFR} & {ROC} \\ \midrule
        PrimeVul & 11.7\% & 31.3\% & 41.4\% & 25.0 & 0.18 & 61.8\% \\ \midrule
        CVEFixes & 16.4\% & 33.6\% & 47.3\% & 16.7 & 0.22 & 68.6\% \\ \midrule
    \end{tabular}
    \label{Tab:2024cve_generalization} \\
\end{table}

Similar to previous experiments, we evaluate vulnerability prioritization using Top-N accuracies, MFR, and N-MFR. For N-MFR, we apply a normalization factor of 76 LoC, representing the average vulnerable function length in the 2024 CVEFixes dataset. The evaluation covers 147 vulnerable functions from 100 CVEs. As shown in Table~\ref{Tab:2024cve_generalization}, \tool{} demonstrates similar Top-1 to Top-5 accuracies across both datasets, with slightly better performance on CVEFixes. This improvement is expected, as the shorter average function length in CVEFixes theoretically results in higher prioritization metrics. After normalization, the N-MFR scores converge to 0.18 and 0.22 for PrimeVul and CVEFixes, respectively. This consistency highlights the robustness of \tool{}'s vulnerability prioritization capabilities, even when applied to previously unseen datasets.

For the evaluation of vulnerability classification, we assess all 3,736 functions in the dataset, which consists of 235 vulnerable LoC and over 110k non-vulnerable LoC. As shown in Table~\ref{Tab:2024cve_generalization}, \tool{} achieves a higher ROC-AUC score of 68.6\%
($p<0.0001$)
on the 2024 CVEFixes dataset compared to 61.8\% on the PrimeVul dataset. This result confirms that our anomaly-based approach generalizes and is robust to potential data leakage, accurately detecting vulnerabilities in unseen datasets.

\section{\!\!\!Real-World Integration:\! Enhancing Fuzzing with \tool{}}
In Section~\ref{sec:eval}, \tool{} demonstrated superior vulnerability detection and prioritization capabilities compared to existing ML-based detectors. However, due to the inherent difficulty of vulnerability review, even a low false-positive rate can lead to hours of manual investigation, making ML-based detectors, including \tool{}, potentially impractical as standalone static analysis tools in production settings. Nonetheless, we hypothesize that \tool{} can still be effective, especially when used in combination with other vulnerability analysis tools, e.g., with fuzzers.

One promising application of \tool{} is to guide fuzzing seed selection. Traditional fuzzers use initial inputs or \emph{seeds} to generate large numbers of mutated inputs in each iteration, aiming to maximize code coverage. However, many of these seeds, despite increasing coverage, may not exercise vulnerable code paths. This leads fuzzers to waste effort on ``uninformative'' inputs. In contrast, anomaly scores produced by \tool{} can help prioritize seeds that are more likely to interact with potentially vulnerable code regions, thereby accelerating bug discovery.

As a proof of concept, we integrated \tool{} with three fuzzers included in the Magma benchmark~\cite{hazimeh_magma_2020}: AFL++\cite{AFLplusplus-Woot20}, Honggfuzz\cite{googlehonggfuzz_2025}, and LibFuzzer~\cite{libfuzzer_nodate}. The Magma benchmark includes 24 binaries across 9 fuzzing targets, and provides an initial corpus of 19,748 seeds. For each seed, we executed the binary to determine its code coverage. We then selected the top 10\% most anomalous lines based on \tool{}’s scoring results, and retained only those seeds that covered at least one of these lines. This filtering process removed 34\% of seeds, leaving 13,059 prioritized ones.

\begin{figure}[h!]
    \centering
    \includegraphics[width=0.68\linewidth]{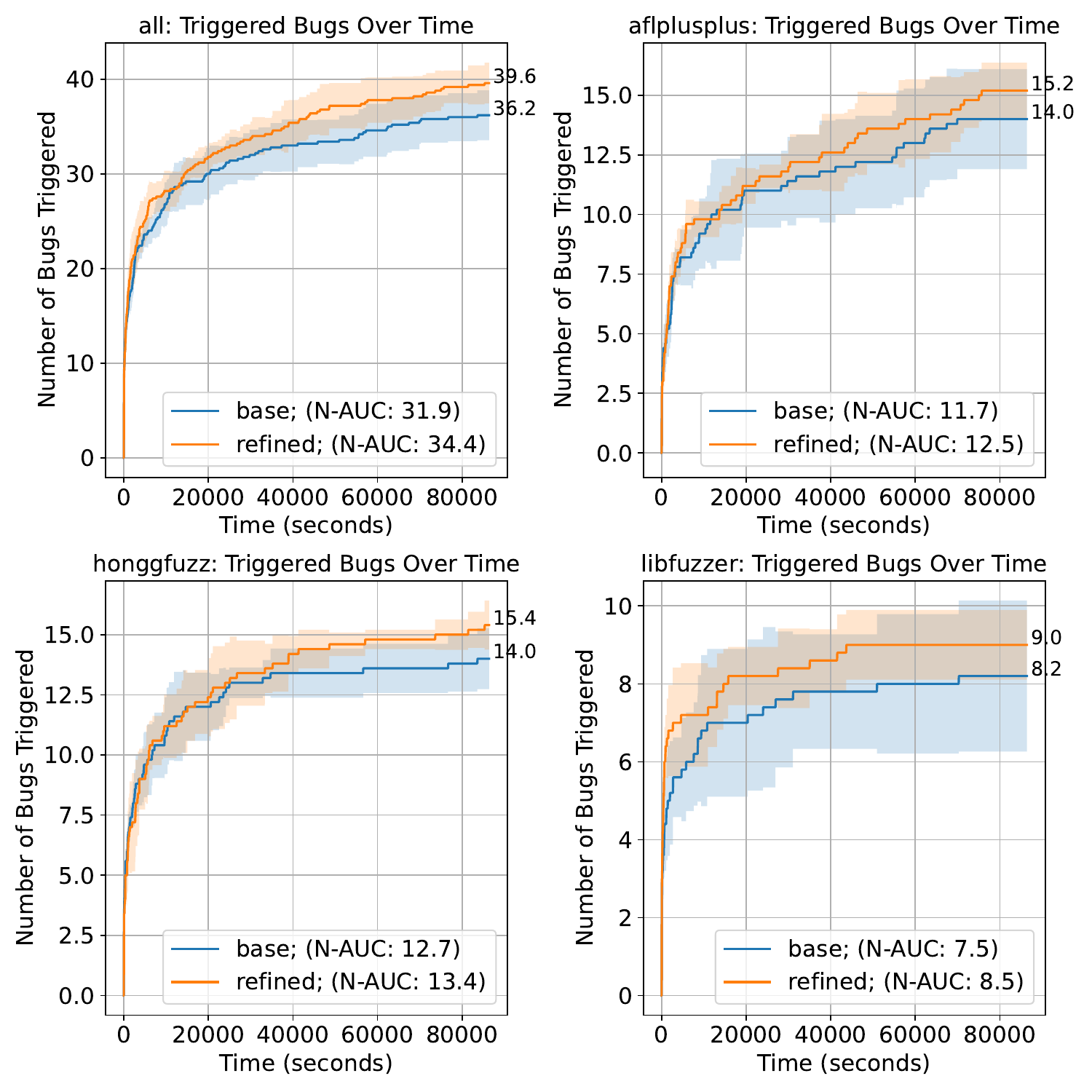}
    \Description[Triggered bugs over time for original and ANVIL-refined fuzzing seeds.]{
    The figure contains four time-series plots showing cumulative bugs triggered over 24 hours for all fuzzers combined and separately for AFL++, Honggfuzz, and LibFuzzer. In each plot, the ANVIL-refined seed set triggers more bugs and has higher normalized area under the curve than the original seed set. Across all fuzzers, refined seeds reach 39.6 bugs versus 36.2 for original seeds.
    }
    \caption{Number of bugs triggered over time, filled with standard deviations}
    \label{fig:magma_all}
\end{figure}

We then conducted 24-hour fuzzing campaigns using both the original and the refined seed sets, repeating each run five times per fuzzer, on a 128-core Intel Xeon Gold 6530 server. Figure~\ref{fig:magma_all} shows the cumulative number of bugs triggered over time. After 24 hours, the seed set refined by \tool{} enabled fuzzers to discover 9.4\% more bugs (39.6 vs. 36.2). We also report Normalized Area Under the Curve (N-AUC), which accounts for both the number of bugs found and the time taken to trigger them. N-AUC improved by 7.8\%, showing that integrating \tool{} with fuzzers not only improves bug yield but also increases efficiency.



\subsubsection{Zero-day Vulnerabilities.}
We applied \tool{} in combination with LibFuzzer and discovered two zero-day vulnerabilities in SFML\footnote{\url{https://github.com/SFML/SFML}}, a widely used multimedia library with over 11{,}000 GitHub stars. SFML was selected from approximately ten popular C/C++ libraries after preliminary screening for reproducible fuzzing and in-depth analysis suitability. The two bugs are categorized as a heap-use-after-free and a heap-buffer-overflow, respectively. 

The heap-use-after-free vulnerability was caused by missing synchronization between SFML's internal audio playback thread and the main thread when stopping a specially crafted audio track prioritized by \tool{}. We reported this bug as a GitHub issue\footnote{\url{https://github.com/SFML/SFML/issues/3503}}, 
later patched by the developers.


The heap-buffer-overflow vulnerability was responsibly disclosed to MITRE and assigned the reserved identifier CVE-2025-50940. The bug resides in the \texttt{minimp3}~\cite{lion_lieffminimp3_2026} library used by SFML, and is caused by a mismatch between a pointer and its associated size, leading to an out-of-bounds memory access. The issue was then fixed by the SFML maintainers by replacing the affected library\footnote{\url{https://github.com/SFML/SFML/pull/2995}}.

The two vulnerabilities span different temporal relationships with the LLM’s knowledge cutoff: the heap-buffer-overflow originates from pre-cutoff code, while the heap-use-after-free involves post-cutoff modifications, demonstrating that \tool{} does not rely on memorization or training-data leakage.
We also checked SFML using CodeQL~\cite{codeql_nodate}, a SOTA static analyzer. When using its ``extended'' query suite (known for high false-positive rates), CodeQL flagged 48 potential issues. However, none were related to the two zero-days. In contrast, \tool{} was able to prioritize the root-cause lines of both vulnerabilities within the top 10\% most anomalous lines, effectively guiding LibFuzzer to trigger the bugs.
\section{Discussion}

\tool{} demonstrates effective vulnerability detection by reframing the task as anomaly detection. From a systems perspective, \tool{} is intentionally designed to operate directly on raw source code and to be fully parallelizable, enabling independent LLM inference across lines at scale. This simplicity facilitates deployment while leaving room for refinement. For example, lightweight static analysis or preprocessing may further enhance anomaly discrimination by providing richer semantic signals, such as LLM-generated summaries or inter-function-level information injected as auxiliary context. At the same time, such techniques introduce additional overhead and design trade-offs that require careful evaluation. Exploring these extensions forms a natural part of our roadmap.

Beyond this, our current fuzzing integration shows that anomaly scores effectively prioritize vulnerability-prone regions, motivating further exploration of anomaly-guided dynamic analysis. To our knowledge, while prior work has used ML models to select seeds for optimizing coverage~\cite{259709,10.1145/3664603}, prioritizing seeds based on bug likelihood remains unexplored. While our current integration focuses on filtering the initial seed corpus, richer interactions with fuzzing workflows are possible. As fuzzing progresses, newly generated seeds may exercise highly anomalous code regions, suggesting that dynamically incorporating anomaly feedback could further improve performance. Moreover, anomaly scores naturally lend themselves to guiding optimization-based directed fuzzers such as AFLGo~\cite{bohme_directed_2017}, where anomaly-ranked locations could serve as fuzzing targets. Realizing such integrations, however, requires careful consideration of how targets are selected and grouped, as anomaly-based scanning may identify many locations across diverse execution paths. Designing strategies that balance focus and coverage remains an important direction for future work.
\section{Related Work}

\subsubsection{Anomaly Detection.}
Prior work has leveraged anomalous behaviour to identify bugs. Yun et al.~\cite{yun_apisan_2016} detect API misuse by learning correct usage patterns, while Ahmadi et al.~\cite{ahmadi_finding_2021} identify bugs by clustering inconsistent code within a project. To our knowledge, the only prior LLM-based anomaly-guided vulnerability detector compares LLM outputs to ground truth using string distance~\cite{ahmad_flag_2023}, which lacks contextual understanding and relies on pre-existing comments. In contrast, \tool{} incorporates context-aware, multi-faceted anomaly scoring without assuming comment availability.

\subsubsection{ML-Based Vulnerability Detection.}
Prior work has explored CNNs, GNNs, and LSTM-based models for vulnerability detection~\cite{russell_automated_2018,cheng_deepwukong_2021,li_vuldeepecker_2018}, but such approaches often fail to generalize to unseen real-world bugs~\cite{chakraborty_deep_2022}. Other work considers alternative code representations, such as code slices and bug-triggering paths~\cite{mirsky_vulchecker_2023,cheng_how_2022}. Machine learning has also been used to enhance fuzzing, including predicting inputs for smart contracts~\cite{he_learning_2019}, enabling input-format-aware fuzzing via basic-block classification~\cite{shi_aifore_2023}, and deriving protocol grammars using LLMs~\cite{meng_large_2024}.

\subsubsection{LLM-Based Vulnerability Detection.}
Recent work explores using LLMs for vulnerability detection. Zhang et al.~\cite{10.1145/3639478.3643065} show that prompt-based ChatGPT usage can outperform traditional detectors, while Lin et al.~\cite{lin_large_2025} systematically evaluate open-source LLMs for file-level detection across model sizes, quantization settings, and context lengths. However, these approaches typically rely on fixed or maximum-length contexts and carefully crafted prompts. Wu et al.~\cite{10.1145/3691620.3695013} leverage in-context learning and semantic similarity for function localization given known CVE descriptions, targeting localization rather than detection of unknown vulnerabilities. In contrast, \tool{} dynamically adjusts context size, requires no vulnerability descriptions, and exploits the FIM pretraining objective without task-specific prompt engineering.

Several studies combine LLMs with structured program representations. Li et al.~\cite{li-etal-2025-clever} apply contrastive learning over code property graphs and vulnerability descriptions to learn vulnerability-aware representations, while Wen et al.~\cite{wen_scale_2024} and Ni et al.~\cite{10.1145/3731557} integrate LLM-derived features with program structures such as ASTs, CFGs, and PDGs for supervised vulnerability detection. Lekssays et al.~\cite{10.5555/3766078.3766104} further extract vulnerability-related execution paths via an LLM-based graph query generator and augment them into code slices for classification.

These approaches incur nontrivial overhead due to program graph construction and typically rely on labelled data or auxiliary information such as vulnerability descriptions or execution paths. In contrast, \tool{} operates directly on raw source code, is fully parallelizable, and requires no labelled vulnerability data. Finally, it is worth noting that all LLM-based approaches discussed in this subsection evaluate vulnerability detection at coarse granularities, including function-, file-, or project-level detection, whereas \tool{} performs line-level detection, providing finer-grained and more actionable vulnerability localization.
\section{Conclusion}

To address the limitations of supervised-learning-based vulnerability detectors, we propose \tool{}, which leverages LLM mispredictions to guide vulnerability detection. 
Our method treats discrepancies between LLM-predicted and actual code as anomalies that can signal vulnerabilities. We demonstrate that Adaptive Context (AC) outperforms fixed-sized contexts for LLM predictions. Our approach generalizes across LLM sizes, architectures, and datasets, including vulnerabilities disclosed after model training. In comparative evaluations, \tool{} outperforms state-of-the-art supervised models, LineVul, LineVD, and LLMAO, despite requiring no labelled training data. Finally, by integrating \tool{} with fuzzers, we discover two previously unknown vulnerabilities in popular software.
\subsubsection{Acknowledgments}
This research was supported in part by NSERC Discovery Grant RGPIN-2018-05931 and NSERC-CSE Research Communities Grant ALLRP 588144-23. David Lie is supported by Tier 1 Canada Research Chair CRC-2019-00242. Researchers funded through the NSERC-CSE Research Communities Grants do not represent the Communications Security Establishment Canada or the Government of Canada. Any research, opinions or positions they produce as part of this initiative do not represent the official views of the Government of Canada.

\begin{subappendices}
    \renewcommand{\thesection}{\Alph{section}}

\section{Model Architecture and Model Size}\label{sec:model_arch_size}
This appendix evaluates whether anomaly-based vulnerability detection generalizes across different LLM architectures and model sizes. We perform the FIM workload on multiple open-source code LLMs that support FIM, selected from the top-ranked base models on Huggingface’s Big Code Models Leaderboard as of May~1,~2025~\cite{noauthor_big_nodate}, and additionally study the effect of model scaling by comparing different sizes within the same architecture.


Table~\ref{Tab:model_list} summarizes the evaluated models, covering three architectures and sizes ranging from 0.5B to 33B parameters. For each model, we apply AC and compute ROC-AUC using the $\delta_{hybrid}$ anomaly score. As shown in Table~\ref{Tab:different_llm}, all models achieve ROC-AUC above 60\% ($p<0.0001$). Pairwise comparisons further show consistent anomaly rankings across models, with Spearman rank correlation exceeding 0.6 for all pairs and Cohen’s $d$ below 0.2 in 53 of 55 comparisons, indicating that the anomaly-based detection approach generalizes well across model architectures. Across model sizes within the same architecture, larger models generally achieve higher ROC-AUC, suggesting that discriminative capability improves with neural model scaling.

\begin{table}[ht!]
\centering
\Description[Code LLM architectures used in experiments.]{
The table lists the code LLM families evaluated in the experiments. CodeLlama includes 7B and 13B models trained on about 29 billion lines of code. DeepSeek-Coder includes 1.3B, 6.7B, and 33B models trained on about 114 billion lines of code. Qwen2.5-Coder includes 0.5B, 1.5B, 3B, 7B, 14B, and 32B models trained on about 314 billion lines of code.
}
\caption{Code LLMs used in the experiments}
 \begin{tabular}{l l l l} 
 \hline
 Model Name &  Sizes & Training data~\footnotemark & Publication\\ [0.5ex] 
 \hline
    CodeLlama & 7B, 13B & 29B LOC & Aug 2023\\
    DeepSeek-Coder & 1.3B, 6.7B, 33B & 114B LOC & Jan 2024\\ 
    Qwen2.5-Coder & 0.5B, 1.5B, 3B,7B, 14B, 32B & 314B LOC & Nov 2024\\[0.5ex] 
 \hline
 \label{Tab:model_list}
 \end{tabular}
\end{table}\footnotetext{Approximated using 4 characters per token and 70 characters per LOC.}

\begin{table}[h!]
\centering
\Description[ROC-AUC across model families and sizes on Magma.]{
The table reports ROC-AUC values for Qwen, DeepSeek, and CodeLlama models of different sizes on Magma. All reported values exceed 60 percent. Qwen ranges from 61.0 percent at 0.5B to 65.6 percent at 32B, DeepSeek ranges from 60.5 percent to 62.3 percent, and CodeLlama reaches 63.9 percent at 7B and 66.3 percent at 14B.
}
\caption{ROC-AUC of different LLMs on the Magma Dataset}
    \begin{tabular}{ |c|c|c|c|c|c|c| }
         \hline
         & \multicolumn{6}{c|}{Model Size} \\
         \hline
         LLM & 0.5B & 1.5B & 3B & 7B & 14B & 32B \\
         \hline
         Qwen & 61.0\% & 61.4\% & 65.3\% & 64.0\% & 64.1\% & 65.6\% \\ \hline
         DeepSeek & N/A & 60.5\% & N/A & 61.5\% & N/A & 62.3\% \\ \hline
         CodeLlama & N/A & N/A & N/A & 63.9\% & 66.3\% & N/A \\
         \hline
    \end{tabular}
    \label{Tab:different_llm}
\end{table} 

\section{Precision, Recall and F1 Score}\label{sec:rq3_PRF1}

Table~\ref{Tab:primevul_prf_cmp} reports Precision, Recall, and F1 scores for \tool{} and baselines on the full PrimeVul dataset (33,820 non-vulnerable vs.\ 924 vulnerable lines) and a balanced subset (924/924). For each method, the decision threshold is chosen to maximize F1, and the corresponding Precision and Recall are reported.

\tool{} achieves the highest F1 and Precision in both settings. While some baselines attain higher Recall at their F1-optimal thresholds, this comes at the cost of substantially lower Precision due to increased false positives, resulting in lower overall F1 scores than \tool{}. Because Precision and Recall are reported at method-specific operating points and reflect different trade-offs, we use ROC-AUC as the primary metric in the main text, and include Precision, Recall, and F1 results in the appendix for completeness and consistency with prior work.

\begin{table}[h!]
\centering
\caption{Precision, Recall and F1 Scores on PrimeVul Dataset}
\Description[Precision, recall, and F1 scores on PrimeVul.]{
The table reports precision, recall, and F1 for ANVIL and three baselines on the full and balanced PrimeVul datasets. ANVIL has the best F1 and precision in both settings: 1.19 percent F1 on the full dataset and 67.56 percent F1 on the balanced dataset. LLMAO, LineVD, and LineVul achieve higher recall in some settings but lower precision and lower F1.
}
\begin{tabular}{@{\hspace{0.5em}}l@{\hspace{0.5em}}ccc@{\hspace{1.2em}}ccc@{\hspace{0.5em}}} \toprule
\multirow{2}{*}{Method} &
\multicolumn{3}{c}{Full Dataset} &
\multicolumn{3}{c}{Balanced Dataset} \\
\cmidrule(lr){2-4}\cmidrule(lr){5-7}
& Prec & Recall & F1$\uparrow$
& Prec & Recall & F1$\uparrow$ \\
\midrule
ANVIL   & \textbf{0.62\%} & 15.60\% & \textbf{1.19\%} & \textbf{52.75\%} & 93.95\%  & \textbf{67.56\%} \\
LLMAO   & 0.43\% & \textbf{77.95\%} & 0.85\% & 51.01\% & 97.60\% & 67.00\% \\
LineVD  & 0.24\% & 21.95\% & 0.47\% & 50.00\% & \textbf{100.00\%} & 66.67\% \\
LineVul & 0.33\% & 10.77\% & 0.64\% & 50.00\% & \textbf{100.00\%} & 66.67\% \\
\bottomrule
\end{tabular}
\label{Tab:primevul_prf_cmp}
\end{table}

\section{Ablation Study}\label{sec:ablation}
While RQ1 (Section~\ref{sec:eval_anom_score}) evaluates anomaly score components in isolation, this ablation study assesses their contributions within the combined hybrid score $\delta_{hybrid}$ by selectively removing components. As shown in Table~\ref{Tab:ablation}, the full hybrid score consistently achieves the best overall performance across metrics (with $p<0.0001$ for ROC-AUC), and performance degrades as components are removed.

The only exception is Top-1 accuracy, which exhibits higher variance and where $\delta_{hybrid}$ narrowly misses the best result (11.7\%). This behaviour arises because the model often assigns high anomaly scores to lines adjacent to the vulnerable line, occasionally ranking them above the exact root-cause line. Overall, the hybrid scoring function remains the most robust, with each component contributing to effective vulnerability detection.
\begin{table}[h!]
\centering
\Description[Ablation results for hybrid anomaly-score components.]{
The table evaluates the hybrid anomaly score after removing components. Using all four components, loss, exact match, variance, and AST complexity, gives the best ROC, Top-3, Top-5, and normalized MFR values: 61.8 percent, 31.3 percent, 41.4 percent, and 0.18. Removing components progressively lowers most metrics, although the loss plus exact-match variant has the highest Top-1 value at 12.1 percent.
}
\caption{Ablation Study}
    \setlength{\tabcolsep}{4.5pt}
    \renewcommand{\arraystretch}{1.2}
    \begin{tabular}{ |c|c|c|c|c|c|c|c|c| }
         \hline
         loss & EM &  var & ast & ROC$\uparrow$ & Top-1$\uparrow$ & Top-3$\uparrow$ & Top-5$\uparrow$ & N-MFR$\downarrow$ \\
         \hline
         \checkmark & \checkmark & \checkmark & \checkmark & \textbf{61.8\%} & 11.7\%          & \textbf{31.3\%} & \textbf{41.4\%} & \textbf{0.18}  \\
         \checkmark & \checkmark & \checkmark &            & 59.0\%          & 10.2\%          & 28.5\%          & 41.0\%          & 0.20 \\
         \checkmark & \checkmark &            &            & 57.8\%          & \textbf{12.1\%} & 27.3\%          & 39.8\%          & 0.20 \\ 
         \checkmark &            &            &            & 57.1\%          & 11.3\%          & 26.6\%          & 39.1\%          & 0.22  \\
         \hline
    \end{tabular}
    \label{Tab:ablation}
\end{table}
\end{subappendices}

\bibliographystyle{splncs04}
\bibliography{references}
\end{document}